\documentclass[francais]{article-arima}
\usepackage{amssymb,amsbsy,amsmath,amsfonts,amssymb,amscd}

\newtheorem{theoreme}{Th\'eor\`eme}[section]

\newtheorem{proposition}[theoreme]{Proposition}
\newtheorem{corollaire}[theoreme]{Corollaire}

\newtheorem{remarque}{\it Remarque}

\author{Michel FLIESS\protect\footnote{L'auteur d\'{e}die ce travail \`{a} Claude LOBRY, chercheur
aussi brillant qu'original, et ami fid\`{e}le, qui lui a beaucoup
appris, l'analyse non standard par exemple.}}
\title[Rapport signal \`{a} bruit]{Critique du rapport signal \`{a} bruit\\
en communications num\'{e}riques} \subtitle{Questioning the signal to
noise ratio \\ in digital communications}
\address{
INRIA--ALIEN \& LIX (CNRS, UMR 7161) \\
\'Ecole polytechnique \\
91128 Palaiseau \\
France \\
Michel.Fliess@polytechnique.edu }

\resume{On d\'{e}montre que le rapport signal \`{a} bruit, si important en
th\'{e}orie de l'information, devient sans objet pour des communications
num\'{e}riques o\`{u} la d\'{e}modulation s'effectue selon des techniques
nouvelles d'estimation rapide. Calcul op\'{e}rationnel, alg\`{e}bre
diff\'{e}rentielle, alg\`{e}bre non commutative et analyse non standard sont
les principaux outils math\'{e}matiques.}

\abstract{The signal to noise ratio, which plays such an important
r\^{o}le in information theory, is shown to become pointless for digital
communications where the demodulation is achieved via new fast
estimation techniques. Operational calculus, differential algebra,
noncommutative algebra and nonstandard analysis are the main
mathematical tools.} \motscles{Th\'{e}orie de l'information, traitement
du signal, communications num\'{e}riques, porteuses, symboles,
modulation, d\'{e}modulation, bruits, estimation, rapport signal \`{a}
bruit, calcul op\'{e}rationnel, alg\`{e}bre diff\'{e}rentielle, alg\`{e}bre non
commutative, analyse non standard.}  \journal[2007 International Conference in
Honor of Claude Lobry]{\textbf{\arima}}{9}{1}{2008}{419}{429}

\setboolean{isrubrique}{true}
\begin{document}
\maketitlepage
\pagestyle{otherpage}

\selectlanguage{english}
\section*{Extended english abstract}

\noindent{\bf Introduction}

\noindent The symbol to be transmitted is modulating a carrier,
which is assumed to be a solution of a linear differential equation
with polynomial coefficients. Most signals utilized in practice,
like $\sum_{\tiny{\mbox{\rm finite}}} A_\iota \sin (\omega_\iota t +
\varphi_\iota)$, $A_\iota, \omega_\iota, \varphi_\iota \in
\mathbb{R}$, ${\mbox {\rm sinc}} ~t = \frac{\sin t}{t}$, $\frac{\cos
t}{1 + t^2}$, do satisfy this property. New algebraic estimation
techniques \cite{esaim,garnier} permit to achieve demodulation even
with very ``strong'' corrupting additive noises and therefore to
question the importance of the signal to noise ratio, which is
playing such a crucial r\^ole in information theory (see
\cite{bell,ire} and \cite{battail,blahut,brillouin,cover,proakis}).
Operational calculus, differential algebra and nonstandard analysis
are the main mathematical tools.
\\

\noindent{\bf Identifiability}

\noindent Let $k_0 (\Theta)$ be the field generated by a finite set
$\Theta = \{\theta_1, \dots, \theta_\varrho \}$ of unknown {\em
parameters}, where $k_0$ is a ground field of characteristic $0$. We
are utilizing the classical notations of operational calculus
\cite{yosida}. Introduce the differential field
\cite{chambert,singer} $\bar{k}(s)$ of rational functions in the
indeterminate $s$ over the algebraic closure $\bar{k}$ of $k_0
(\Theta)$, with derivation $\frac{d}{ds}$. Any {\em signal} $x$, $x
\not\equiv 0$, is assumed to satisfy a homogeneous linear
differential equation with coefficients in $\bar{k}(s)$ and
therefore to belong to a Picard-Vessiot extension
\cite{chambert,singer} of $\bar{k}(s)$. Write $\bar{k}(s)
[\frac{d}{ds}]$ the noncommutative ring of linear differential
operators with coefficients in $\bar{k}(s)$. The left $\bar{k}(s)
[\frac{d}{ds}]$-module spanned by $x$ and $1$ is a
$\bar{k}(s)$-vector space of finite dimension $n + 1$, $n \geq 0$.
It yields the {\em minimal} non necessarily homogeneous linear
differential equation (\ref{nh}) where the polynomials $p, q_0,
\dots, q_{n}$ in $\bar{k}[s]$ are coprime. Introduce the square
matrix $\mathfrak{M}$ of order $N + M + 1$, where the
$\xi^{\tiny\mbox{\rm th}}$ line, $0 \leq \xi \leq N + M$, is
(\ref{line}). Techniques stemming from Wronskian determinants
\cite{chambert,singer} demonstrate that the rank of $\mathfrak{M}$
is $N + M$. It follows that the coefficients of the polynomials $p,
q_0, \dots, q_{n}$ are {\em projectively linearly identifiable}
\cite{esaim,garnier}. Consider as a particular case $x =
\frac{p(s)}{q(s)}$ where the polynomials $p, q \in \bar{k} [s]$ are
coprime. Then the coefficients of $p$ and $q$ are also projectively
linearly identifiable.
\\

\noindent{\bf Perturbations and estimators}

\noindent Assume that the unknown parameters $\Theta$ are {\em
linearly identifiable} \cite{esaim,garnier}. With and additive
perturbation $w$, we obtain the estimator (\ref{estim}) where
$\mathfrak{A}$, and $\mathfrak{B}$, $\mathfrak{C}$ are respectvely
$\varrho \times \varrho$ and $\varrho \times 1$ matrices, such that
the entries of $\mathfrak{A}$ and $\mathfrak{B}$ belong to
$\mbox{\rm span}_{k_0 (s) [\frac{d}{ds}]} (1, x)$ and those of
$\mathfrak{C}$ to $\mbox{\rm span}_{R [\frac{d}{ds}]} (w)$, where
$R$ is the localized ring \cite{lang} $k_0(\Theta)[s] (k_0
[s])^{-1}$. Moreover $\det ( \mathfrak{A}) \neq 0$. It is always
possible to obtain an estimator which is {\em strictly polynomial}
with respect to $\frac{1}{s}$, i.e., where the rational functions in
$s$ are polynomials in $\frac{1}{s}$ without constant terms. As in
\cite{ans} it yields in the time domain, if $x$ is an analytic
function, Formula (\ref{estimat}) where $c$ is a constant, $[0, t]$
is the {\em estimation time window}, the {\em divisor} $\delta (t)$
is an analytic function such that $\delta (0) = 0$, $[ \theta_\iota
]_e (t)$ is the estimated value of $\theta_\iota$ at time $t$.
\\

\noindent{\bf Noises}

\noindent We are considering two types of perturbations, which are
{\em noises} in the sense of \cite{ans}:
\begin{itemize}
\item The first noise, which is zero-mean, is a
finite sum $\sum_{\tiny{\mbox{\rm finite}}} A_i \sin (\Omega_i t +
\varphi_i)$ where the frequencies $\Omega_i
>0$ are unlimited.

\item Let $^*\mathbb{N}$, $^*\mathbb{R}$ be the nonstandard
extensions \cite{robinson} of $\mathbb{N}$, $\mathbb{R}$. Replace
$[0, 1] \subset {\mathbb{R}}$ by the hyperfinite set \cite{robinson}
${\mathrm{I}} = \{0, \frac{1}{\bar{N}}, \dots, \frac{\bar{N} -
1}{\bar{N}}, 1 \}$, where $\bar{N} \in {^*\mathbb{N}}$ is unlimited.
A {\em zero-mean white noise} is a function $w: {\mathrm{I}}
\rightarrow {^*\mathbb{R}}$, $\iota \mapsto w(\iota) = A n(\iota)$,
where
\begin{itemize}
\item $A \in {^*\mathbb{R}}$ is a constant, such that
$\frac{A^2}{\bar{N}}$ is limited,
\item the $n(\iota)$ are independent
zero-mean random variables with a normalized covariance $1$.
\end{itemize}
\end{itemize}
The estimator (\ref{estimat}) yields ``good'' values for the unknown
parameters for any limited values of the amplitudes $A_i$, $A$, and
even for some unlimited values of them.

\newpage

\selectlanguage{francais}

\section{Introduction}
\label{intro} Le {\em rapport signal \`{a} bruit}\footnote{Rappelons au
lecteur peu au fait que toute transmission physique de signal est
perturb\'{e}e, no\~tamment par du \og bruit \fg. L'extraction des
informations utiles malgr\'{e} ces alt\'{e}rations est un but essentiel en
traitement du signal et en th\'{e}orie des communications. Que l'on
pense par exemple aux {\em codes correcteurs d'erreurs}, o\`{u}, comme
les math\'{e}maticiens le savent, la th\'{e}orie des {\em codes en blocs}
est d'une grande richesse alg\'{e}brique.}, que l'on retrouve dans les
formules de la th\'{e}orie de l'information, telle qu'elle s'est impos\'{e}e
depuis Shannon (voir \cite{bell,ire} et, par exemple, dans la vaste
litt\'{e}rature sur le sujet,
\cite{battail,blahut,brillouin,cover,proakis}), est un ingr\'{e}dient
fondamental pour d\'{e}finir la qualit\'{e} des communications. Le but de ce
travail\footnote{Voir \cite{arxiv} pour une version pr\'{e}liminaire.}
est de d\'{e}montrer qu'une nouvelle approche de l'estimation rapide et
du bruit (voir \cite{ans} et sa bibliographie) rend ce rapport sans
objet dans un certain cadre num\'{e}rique. Revoyons, donc, le \og
paradigme de Shannon \fg. Le {\em symbole} \`{a} transmettre (voir, par
exemple, \cite{glavieux,proakis}) {\em module} une {\em porteuse}
$z(t)$, solution d'une \'{e}quation diff\'{e}rentielle lin\'{e}aire \`{a}
coefficients polynomiaux:
$$
\sum_{\tiny{\mbox{\rm finie}}} a_\nu (t)
z^{(\nu)} (t) = 0, \quad \quad ~ ~ a_\nu \in \mathbb{C}[t]
$$
La plupart des signaux utilis\'{e}s en pratique, comme une somme
trigonom\'{e}trique finie $\sum_{\tiny{\mbox{\rm finie}}} A_\iota \sin
(\omega_\iota t + \varphi_\iota)$, un sinus cardinal $\frac{\sin
(\omega t)}{t}$ ou un cosinus sur\'{e}lev\'{e} $\frac{\cos (\omega t)}{1 +
t^2}$, $A_\iota$, $\omega_\iota$, $\varphi_\iota$, $\omega \in
\mathbb{R}$, v\'{e}rifient une telle \'{e}quation, qui se traduit dans le
domaine op\'{e}rationnel (cf. \cite{yosida}), pour $t \geq 0$, par
\begin{equation}\label{nh0} \sum_{\tiny{\mbox{\rm finie}}} a_\nu (-
\frac{d}{ds}) s^\nu \hat{z} = I(s) \end{equation} o\`{u} $I \in
\mathbb{C}[s]$ est un polyn\^{o}me dont les coefficients d\'{e}pendent des
conditions initiales en $t = 0$. La {\em d\'{e}modulation} revient,
alors, \`{a} estimer certains des coefficients de (\ref{nh0}). On y
parvient, ici, gr\^{a}ce \`{a} des techniques alg\'{e}briques r\'{e}centes (cf.
\cite{esaim,garnier}).

Un bruit, selon \cite{ans}, est une fluctuation rapide, que l'on
d\'{e}finit de fa\c{c}on efficace et \'{e}l\'{e}gante gr\^{a}ce \`{a} l'analyse non
standard\footnote{Voir aussi \cite{lobry}.}. Les calculs du {\S}
\ref{bruit} sont effectu\'{e}s avec une somme finie de sinuso\"{\i}des \`{a} tr\`{e}s
hautes fr\'{e}quences et un bruit blanc, dont la d\'{e}finition non standard
clarifie l'approche usuelle des manuels de traitement du signal. Ils
d\'{e}montrent la possibilit\'{e} d'obtenir de \og bonnes \fg ~ estimations
avec des bruits \og tr\`{e}s forts \fg, c'est-\`{a}-dire de \og grandes \fg
~ puissances, fait confirm\'{e} par des simulations num\'{e}riques et des
expriences de laboratoire (voir
\cite{fmmsr,liu,mboup,ajaccio,neves,trapero,trapero-bis,trapero-ter}).
Les imperfections, in\'{e}vitables en pratique, proviennent de
l'implantation num\'{e}rique des calculs, notamment de celui des
int\'{e}grales (voir \cite{liu,mboup}), des interf\'{e}rences entre symboles
(voir \cite{battail,glavieux,proakis0,proakis} et leurs
bibliographies), et du fait que les bruits ne sont pas
n\'{e}cessairement centr\'{e}s (voir \`{a} ce propos le {\S} 3.2.2 de \cite{ans}).

Calcul op\'{e}rationnel et alg\`{e}bre diff\'{e}rentielle aux {\S} \ref{algebre} et
\ref{perturb}, analyse non standard au {\S} \ref{bruit} sont les
principaux outils math\'{e}matiques.

\begin{remarque}
Ajoutons pour le lecteur \'{e}tranger, voire hostile, \`{a} l'analyse non
standard\footnote{Lobry \cite{N} a \'{e}crit un pamphlet \'{e}difiant sur
l'histoire \og agit\'{e}e \fg ~ et la r\'{e}ception \og houleuse \fg ~ de
cette analyse, en d\'{e}pit (\`{a} cause?) de sa beaut\'{e} et de sa puissance
inconstestables. Les propres d\'{e}boires de l'auteur lui ont prouv\'{e} que
ce t\'{e}moignage n'est pas exag\'{e}r\'{e}!} qu'il est loisible de la remplacer
par des consid\'{e}rations \og classiques \fg. On y perdrait en
concision et, \`{a} notre avis, en intuition.
\end{remarque}

\begin{remarque}
Avec des signaux analytiques par morceaux (le sens du mot {\em
analytique} est celui de la th\'{e}orie des fonctions et non pas, ici,
celui usuel en traitement du signal (cf.
\cite{battail,proakis0,proakis})), qui ne satisfont pas d'\'{e}quations
diff\'{e}rentielles connues \`{a} l'avance, on utilise, selon les m\^{e}mes
principes alg\'{e}briques, des d\'{e}rivateurs num\'{e}riques \`{a} fen\^{e}tres
glissantes pour obtenir les estimations (voir \cite{ath} et
\cite{compression,nl,join}, leurs exemples et leurs bibliographies).
On ne peut, alors, esp\'{e}rer les m\^{e}mes r\'{e}sultats que pr\'{e}c\'{e}demment.
\end{remarque}

\begin{remarque}
La possibilit\'{e} de liens entre th\'{e}orie de l'information et m\'{e}canique
quantique a \'{e}t\'{e} examin\'{e}e par divers auteurs (voir, par exemple,
\cite{brillouin,austria,green}). Rappelons \`{a} ce propos que
l'approche du bruit en \cite{ans} a d\'{e}j\`{a} conduit \`{a} une tentative
nouvelle de formalisation du quantique \cite{mecaqua}, qui sera
compl\'{e}t\'{e}e gr\^{a}ce \`{a} un r\'{e}sultat remarquable et tout r\'{e}cent, d\^{u} \`{a}
Charreton \cite{charreton}.
\end{remarque}

\begin{remarque}
Les techniques d'estimation \'{e}voqu\'{e}es plus haut ont permis des
avanc\'{e}es notables en automatique\footnote{Voir, par exemple,
\cite{linz,diag,nl,esaim,recons,garnier} et leurs bibliographies.
Des questions classiques sur l'identification param\'{e}trique, les
observateurs, le diagnostic et l'att\'{e}nuation de perturbations y
re\c{c}oivent des solutions d'une grande simplicit\'{e} conceptuelle et
faciles \`{a} mettre en {\oe}uvre en temps r\'{e}el. Mentionnons aussi la {\em
commande sans mod\`{e}le} \cite{sm}, particuli\`{e}rement prometteuse.},
lin\'{e}aire ou non.
\end{remarque}

\begin{remarque}
Des travaux en cours portant sur l'ing\'{e}ni\'{e}rie financi\`{e}re devraient
\'{e}galement d\'{e}montrer l'applicabilit\'{e} de notre point de vue \`{a} cette
discipline\footnote{Voir \cite{finance} pour une premi\`{e}re \'{e}bauche.}.
\end{remarque}

\vspace{0.4cm} \noindent{\bf Remerciements}. L'auteur exprime sa
reconnaissace \`{a} O. Gibaru (Lille), M. Mboup (Paris) et \`{a} tous les
membres du projet INRIA--ALIEN pour des \'{e}changes fructueux.

\vspace{0.4cm}

\section{Identifiabilit\'{e}}\label{algebre}
\subsection{\'Equations diff\'{e}rentielles}
Renvoyons \`{a} \cite{chambert} pour des rappels sur les corps,
diff\'{e}rentiels ou non.

Soit $k_0$ le corps de base de caract\'{e}ristique nulle, $\mathbb{Q}$
par exemple. Soit $k_0 (\Theta)$ le corps engendr\'{e} par un ensemble
fini $\Theta = \{\theta_1, \dots, \theta_\varrho \}$ de {\em
param\`{e}tres} inconnus. Soit $\bar{k}$ la cl\^{o}ture alg\'{e}brique de $k_0
(\Theta)$. Introduisons le corps $\bar{k}(s)$ des fractions
rationnelles en l'ind\'{e}termin\'{e}e $s$, que l'on munit d'une structure
de corps diff\'{e}rentiel gr\^{a}ce \`{a} la d\'{e}rivation $\frac{d}{ds}$ (les
\'{e}l\'{e}ments de $k_0$, de $\Theta$ et, donc, de $\bar{k}$, sont des
constantes). Tout {\em signal} $x$, $x \not\equiv 0$, est suppos\'{e}
satisfaire une \'{e}quation diff\'{e}rentielle lin\'{e}aire homog\`{e}ne, \`{a}
coefficients dans $\bar{k}(s)$, et donc appartenir \`{a} une extension
de Picard-Vessiot de $\bar{k}(s)$.

\begin{remarque}
Il suffit pour se convaincre de l'existence d'une telle \'{e}quation
homog\`{e}ne de d\'{e}river les deux membres de (\ref{nh0}) suffisamment de
fois par rapport \`{a} $s$.
\end{remarque}

L'anneau non commutatif $\bar{k}(s) [\frac{d}{ds}]$ des op\'{e}rateurs
diff\'{e}rentiels lin\'{e}aires
$$
\sum_{\tiny{\mbox{\rm finie}}} \varpi_\alpha (s)
\frac{d^\alpha}{ds^\alpha}, \quad \quad ~ \varpi_\alpha (s) \in
\bar{k}(s)
$$
est principal \`{a} droite et \`{a} gauche (cf. \cite{McC}). Le $\bar{k}(s)
[\frac{d}{ds}]$-module \`{a} gauche engendr\'{e} par $x$ et $1$ est un
module de torsion (cf. \cite{McC}), et, donc, un $\bar{k}(s)$-espace
vectoriel de dimension finie, $n + 1$, $n \geq 0$. D'o\`{u} le r\'{e}sultat
suivant qui semble nouveau (cf. \cite{chambert,singer}):

\begin{proposition}
Il existe un entier minimal $n \geq 0$, tel que $x$ satisfait
l'\'{e}quation diff\'{e}rentielle lin\'{e}aire, d'ordre $n$, non n\'{e}cessairement
homog\`{e}ne,
\begin{equation}\label{nh}
\left(\sum_{\iota = 0}^{n} q_\iota \frac{d^\iota}{ds^\iota}\right) x
- p = 0
\end{equation}
o\`{u} les polyn\^{o}mes $p, q_0, \dots, q_{n} \in \bar{k}[s]$ sont premiers
entre eux. Cette \'{e}quation, dite {\em minimale}, est unique \`{a} un
coefficient multiplicatif constant non nul pr\`{e}s.
\end{proposition}

\subsection{Identifiabilit\'{e} lin\'{e}aire projective}
Rappelons que l'ensemble $\Theta = \{\theta_1, \dots, \theta_\varrho
\}$ de param\`{e}tres est dit (cf. \cite{esaim,garnier})
\begin{itemize}
\item {\em lin\'{e}airement identifiable} si, et selement si,
\begin{equation}\label{li}
\mathfrak{A} \left(\begin{array}{c} \theta_1 \\ \vdots \\
\theta_\varrho \end{array} \right) = \mathfrak{B}
\end{equation}
o\`{u}
\begin{itemize}
\item les entr\'{e}es des matrices $\mathfrak{A}$, carr\'{e}e $\varrho
\times \varrho$, et $\mathfrak{B}$, colonne $\varrho \times 1$,
appartiennent \`{a} $\mbox{\rm span}_{k_0 (s) [\frac{d}{ds}]} (1, x)$;
\item $\det ( \mathfrak{A}) \neq 0$.
\end{itemize}
\item {\em projectivement lin\'{e}airement identifiable} si, et
seulement si,
\begin{itemize}
\item il existe un param\`{e}tre, $\theta_1$ par exemple, non nul,
\item l'ensemble $\{ \frac{\theta_2}{\theta_1}, \dots,
\frac{\theta_\varrho}{\theta_1} \}$ est lin\'{e}airement identifiable.
\end{itemize}
\end{itemize}
R\'{e}\'{e}crivons (\ref{nh}) sous la forme suivante:
\begin{equation}\label{nhcoef}
\left( \sum_{\tiny{\mbox{\rm finie}}} a_{\mu \nu} s^\mu
\frac{d^\nu}{ds^\nu} \right) x - \sum_{\tiny{\mbox{\rm finie}}}
b_\kappa s^\kappa = 0
\end{equation}
o\`{u} les $N + 1$ coefficients $a_{\mu \nu}$ et les $M$ coefficients
$b_\kappa$ appartiennent \`{a} $\bar{k}$. La matrice carr\'{e}e
$\mathfrak{M}$ d'ordre $N + M + 1$, dont la $\xi^{\tiny\mbox{\rm
\`{e}me}}$ ligne, $0 \leq \xi \leq N + M$, est
\begin{equation}\label{line} \dots,
\frac{d^\xi}{ds^\xi} \left( s^\mu \frac{d^\nu x}{ds^\nu}\right),
\dots, \frac{d^\xi s^\kappa}{ds^\xi}, \dots
\end{equation}
est singuli\`{e}re d'apr\`{e}s (\ref{nh}) et (\ref{nhcoef}). La minimalit\'{e}
de (\ref{nh}) permet de d\'{e}montrer selon des techniques bien connues
sur le rang du wronskien (cf. \cite{chambert,singer}) que le rang de
$\mathfrak{M}$ est $N + M$. Il en d\'{e}coule:
\begin{theoreme}
Les coefficients $a_{\mu \nu}$ et $b_\kappa$ de (\ref{nhcoef}) sont
projectivement lin\'{e}airement identifiables.
\end{theoreme}
\begin{corollaire}
Posons $x = \frac{p(s)}{q(s)}$, o\`{u} les polyn\^{o}mes $p, q \in \bar{k}
[s]$ sont premiers entre eux. Alors, les coefficients de $p$ et $q$
sont projectivement lin\'{e}airement identifiables.
\end{corollaire}
Il est loisible de supposer l'ensemble des param\`{e}tres inconnus
$\Theta = \{\theta_1, \dots, \theta_\varrho \}$ strictement inclus
dans celui des coefficients $a_{\mu \nu}$ et $b_\kappa$ de
(\ref{nhcoef}), et donc lin\'{e}airement identifiable.

\section{Perturbations et estimateurs}\label{perturb}
Avec une perturbation additive $w$ le capteur fournit non pas $x$
mais $x + w$. Soient
\begin{itemize}
\item $R =  k_0(\Theta)[s] (k_0 [s])^{-1}$ l'anneau {\em localis\'{e}}
(cf. \cite{lang}) des fractions rationnelles \`{a} num\'{e}rateurs dans
$k_0(\Theta)[s]$ et d\'{e}nominteurs dans $k_0 [s]$,

\item $R[\frac{d}{ds}]$ l'anneau non commutatif des op\'{e}rateurs
diff\'{e}rentiels lin\'{e}aires \`{a} coefficients dans $R$.
\end{itemize}
On obtient, \`{a} partir de (\ref{li}), la
\begin{proposition}\label{estimperturb}
Les param\`{e}tres inconnus v\'{e}rifient
\begin{equation}\label{estim}
\mathfrak{A} \left(\begin{array}{c} \theta_1 \\ \vdots \\
\theta_\varrho \end{array} \right) = \mathfrak{B} + \mathfrak{C}
\end{equation}
o\`{u} les entr\'{e}es de $\mathfrak{C}$, matrice colonne $\varrho \times
1$, appartiennent \`{a} $\mbox{\rm span}_{R [\frac{d}{ds}]} (w)$.
\end{proposition}
On appelle (\ref{estim}) un {\em estimateur}. Il est dit {\em
strictement polynomial en $\frac{1}{s}$} si, et seulement si, toutes
les fractions rationnelles en $s$, rencontr\'{e}es dans les coefficients
des matrices $\mathfrak{A}$, $\mathfrak{B}$, $\mathfrak{C}$ de
(\ref{estim}), sont des polyn\^{o}mes en $\frac{1}{s}$ sans termes
constants. On peut toujours s'y ramener en multipliant les deux
membres de (\ref{estim}) par une fraction rationnelle de $k_0(s)$
convenable. On aboutit, alors, dans le domaine temporel, aux
estimateurs consid\'{e}r\'{e}s en \cite{ans}, si l'on suppose l'analyticit\'{e}
du signal:

\begin{equation}\label{estimat}
\delta  (t) \left( [ \theta_\iota ]_e (t) - \theta_\iota \right) =
\sum_{\tiny{\mbox{\rm finie}}} c \int_{0}^{t} \dots
\int_{0}^{\tau_2} \int_{0}^{\tau_1} \tau_{1}^{\nu} w(\tau_1)d\tau_1
d\tau_2 \dots d\tau_k  \quad \quad ~ ~\iota = 1, \dots, \varrho
\end{equation}
o\`u
\begin{itemize}
\item $c$ est une constante,
\item $[0, t]$ est la {\em fen\^etre d'estimation}, de {\em largeur}
$t$,
\item $\delta (t)$ est une fonction analytique,
appel\'{e}e {\em diviseur}, nulle en $0$,
\item $[ \theta ]_e (t)$ est l'estim\'{e}e de $\theta$ en $t$.
\end{itemize}

\section{Bruits}\label{bruit}
Renvoyons \`{a} \cite{robinson} et \cite{diener} pour la terminologie de
l'analyse non standard, d\'{e}j\`{a} utilis\'{e}e en \cite{ans}. Les
propositions \ref{propsin} et \ref{propbr} ci-dessous affinent la
proposition 3.2 de \cite{ans}, o\`{u} les estimations sont obtenues en
temps limit\'{e}, \og court \fg ~ en pratique.

\subsection{Sinuso\"{\i}des hautes fr\'{e}quences}\label{sinus}
La perturbation du {\S} \ref{perturb} est de la forme
$$\sum_{\iota
 = 1}^{M} A_\iota \sin (\Omega_\iota t + \varphi_\iota)$$
o\`{u}
\begin{itemize}
\item $M$ est un entier limit\'{e} standard,
\item les fr\'{e}quences $\Omega_\iota
>0$ sont des constantes illimit\'{e}es,
\item les amplitudes $A_\iota$ sont des constantes, limit\'{e}es ou non,
\item les phases $\varphi_\iota$, $0 \leq \varphi_\iota < 2 \pi$, sont des constantes.
\end{itemize}
Si les quotients $\frac{A_\iota}{\Omega_\iota}$ sont infinit\'{e}simaux,
c'est un bruit centr\'{e}, c'est-\`{a}-dire de moyenne nulle, au sens de
\cite{ans}. Des manipulations \'{e}l\'{e}mentaires des int\'{e}grales it\'{e}r\'{e}es
(\ref{estimat}) conduisent \`{a} la

\begin{proposition}\label{propsin} Si \begin{itemize}
\item les quotients $\frac{A_\iota}{\Omega_\iota}$ sont
infinit\'{e}simaux, et, en particulier, si les $A_\iota$ sont limit\'{e}s,

\item la largeur de la fen\^{e}tre d'estimation est limit\'{e}e et n'appartient
pas au halo d'un z\'{e}ro du diviseur,
\end{itemize}
les estim\'{e}es des param\`{e}tres inconnus, obtenues gr\^{a}ce \`{a}
(\ref{estimat}), appartiennent aux halos de leurs vraies valeurs. Il
n'en va plus de m\^{e}me si l'un des quotients
$\frac{A_\iota}{\Omega_\iota}$ est appr\'{e}ciable.
\end{proposition}
\begin{remarque}
Il existe des valeurs illimit\'{e}es des amplitudes $A_\iota$,
$\sqrt{\Omega_\iota}$ par exemple, telles que les estim\'{e}es
pr\'{e}c\'{e}dentes appartiennent aux halos des vraies valeurs.
\end{remarque}

\subsection{Bruits blancs}
D\'{e}signons par $^*\mathbb{N}$, $^*\mathbb{R}$ les extensions non
standard de $\mathbb{N}$, $\mathbb{R}$. Rempla\c{c}ons l'intervalle $[0,
1] \subset {\mathbb{R}}$ par l'ensemble hyperfini ${\mathrm{I}} =
\{0, \frac{1}{\bar{N}}, \dots, \frac{\bar{N} - 1}{\bar{N}}, 1 \}$,
o\`u $\bar{N} \in {^*\mathbb{N}}$ est illimit\'{e}. Un {\em bruit blanc
centr\'{e}} est une fonction $w: {\mathrm{I}} \rightarrow
{^*\mathbb{R}}$, $\iota \mapsto w(\iota) = A n(\iota)$, o\`{u}
\begin{itemize}
\item l'amplitude $A \in {^*\mathbb{R}}$ est constante,
\item le quotient $\frac{A^2}{\bar{N}}$ est limit\'{e},
\item les $n(\iota)$ sont des variables al\'{e}atoires
r\'{e}elles, suppos\'{e}es centr\'{e}es, de m\^{e}me \'{e}cart-type $1$ normalis\'{e}, et
deux \`{a} deux ind\'{e}pendantes.
\end{itemize}
\begin{remarque}
Cette d\'{e}finition non restreinte au cas gaussien, qu'il convient de
comparer \`{a} celle de \cite{al}, pr\'{e}cise \cite{ans}; elle est inspir\'{e}e
de publications d'ing\'{e}nieurs sur le bruit blanc en temps discret
(voir, par exemple, \cite{proakis0}). Elle clarifie, \`{a} la mani\`{e}re de
\cite{nelson}, l'approche en temps continu usuelle dans les manuels
de traitement du signal (voir, \`{a} ce sujet,
\cite{battail,cover,proakis0,proakis} et leurs bibliographies).
Rappelons que cette approche continue est bas\'{e}e, en g\'{e}n\'{e}ral, sur
l'analyse de Fourier et renvoyons, \`{a} ce sujet, \`{a} \cite{fourier}.
Mentionnons, enfin, les travaux de \cite{gelfand,hida}, bas\'{e}s sur
l'analyse fonctionnelle.
\end{remarque}

\begin{remarque}
Un pas suppl\'{e}mentaire, inutile ici pour nos besoins, consisterait \`{a}
remplacer, comme en \cite{nelson}, les variables al\'{e}atoires
$n(\iota)$ par des analogues \og discrets \fg.
\end{remarque}

Comme au {\S} \ref{sinus}, il vient:

\begin{proposition}\label{propbr} Si
\begin{itemize}
\item le quotient $\frac{A^2}{\bar{N}}$ est
infinit\'{e}simal, et, en particulier, si $A$ est limit\'{e},
\item la largeur $t$, $t \in {\mathrm{I}}$, de la fen\^{e}tre d'estimation
n'appartient pas au halo d'un z\'{e}ro du diviseur,
\end{itemize}
les estim\'{e}es des param\`{e}tres inconnus, obtenues gr\^{a}ce \`{a}
(\ref{estimat}), appartiennent presque s\^{u}rement aux halos de leurs
vraies valeurs. Il n'en va plus de m\^{e}me si le quotient
$\frac{A^2}{\bar{N}}$ est appr\'{e}ciable.
\end{proposition}
\begin{remarque}
Il existe des valeurs illimit\'{e}es de $A$, $\sqrt[3]{\bar{N}}$ par
exemple, telles que les estim\'{e}es pr\'{e}c\'{e}dentes appartiennent presque
s\^{u}rement aux halos des vraies valeurs.
\end{remarque}

\begin{remarque}
Il est loisible de remplacer l'ind\'{e}pendance de $n(\iota)$ et
$n(\iota^\prime)$, $\iota \neq \iota^\prime$, par le fait que
l'esp\'{e}rance du produit $n(\iota) n(\iota^\prime)$ est
infinit\'{e}simale.
\end{remarque}


\begin{thebibliography}{99}

\bibitem[1]{al}\textsc{S. Albeverio, J.E. Fenstad, R. Hoegh-Kr{\o}hn, T. Lindstr{\o}m},
\textit{Nonstandard Methods in Stochastic Analysis and Mathematical
Physics}, Academic Press, 1986.
\bibitem[2]{battail}\textsc{G. Battail}, \textit{Th\'{e}orie de l'information - Application
aux techniques de communication}, Masson, 1997.

\bibitem[3]{blahut}\textsc{R.E. Blahut}, \textit{Principles and
Practice of Information Theory}, Addison-Wesley, 1987.


\bibitem[4]{brillouin}\textsc{L. Brillouin}, \textit{Science and Information Theory}
(2$^{nd}$ ed.), Academic Press, 1962. Traduction fran\c{c}aise de la
1$^{re}$ \'{e}d.: \textit{La science et la th\'{e}orie de l'information},
Masson, 1959.

\bibitem[5]{austria}\textsc{C. Brukner, A. Zeilinger}, \guilo{}Conceptual inadequacy
of the Shannon information in quantum measurements\guilf{},
\textit{Phys. Rev. A}, \volumename\ 63 (2001) 022113.

\bibitem[6]{charreton}\textsc{R. Charreton}, \guilo{}Une loi limite pour les marches al\'{e}atoires
avec des applications physiques\guilf{}, \newblock \textit{C.R.
Acad. Sci. Paris Ser. I}, \volumename\ 345 (2007) 699-703.


\bibitem[7]{cover}\textsc{T.M. Cover, J.A. Thomas}, \textit{Elements of Information Theory},
Wiley, 1991.

\bibitem[8]{chambert} \textsc{A. Chambert-Loir}, \textit{Alg\`{e}bre corporelle}, \'Editions \'Ecole
Polytechnique, 2005. English translation$:$  \textit{A Field Guide
to Algebra}, Springer, 2005.


\bibitem[9]{diener}\textsc{F. Diener, G. Reeb}, \textit{Analyse non standard}, Hermann, 1989.


\bibitem[10]{fourier}\textsc{M. Fliess}, \guilo{}R\'{e}flexions sur la question fr\'{e}quentielle en
traitement du signal\guilf{}, Manuscrit, 2005 (accessible sur {\tt
http$:$//hal.inria.fr/inria-00000461/en/}).

\bibitem[11]{ans}\textsc{M. Fliess}, \guilo{}Analyse non standard du bruit\guilf{},
\newblock \textit{C.R. Acad. Sci. Paris Ser. I}, \volumename\ 342 (2006) 797-802.

\bibitem[12]{mecaqua}\textsc{M. Fliess}, \guilo{}Probabilit\'{e}s et fluctuations quantiques\guilf{},
\newblock \textit{C.R. Acad. Sci. Paris Ser. I}, \volumename\ 344 (2007) 663-668.

\bibitem[13]{arxiv}\textsc{M. Fliess}, \guilo{}Critique du rapport
signal \`{a} bruit en th\'{e}orie de l'information\guilf{}, Manuscrit, 2007
(accessible sur {\tt http$:$//hal.inria.fr/inria-00195987/en/}).

\bibitem[14]{linz} \textsc{M. Fliess, S. Fuchshumer, M. Sch\"{o}berl, K. Schlacher,
H. Sira-Ram\'{\i}rez},
\newblock \guilo{}An introduction to algebraic discrete-time linear parametric identification
with a concrete application\guilf{}, \textit{J. europ. syst.
automat.}, \volumename\ 42 (2008) 211-232.

\bibitem[15]{sm}\textsc{M. Fliess, C. Join}, \guilo{}Commande sans mod\`{e}le et
commande \`{a} mod\`{e}le restreint\guilf{}, \newblock \textit{e-STA},
\volumename\ 5 (2008) (accessible sur {\tt
http$:$//hal.inria.fr/inria-00288107/en/}).

\bibitem[16]{finance}\textsc{M. Fliess, C. Join}, \guilo{}Time series
technical analysis
via new fast estimation methods$:$ A preliminary study in
mathematical finance\guilf{},
\newblock \textit{23$^{rd}$ IAR Workshop Advanced Control Diagnosis -- IAR-ACD08},
Coventry, 2008 (accessible sur {\tt
http$:$//hal.inria.fr/inria-00338099/en/}).


\bibitem[17]{compression}\textsc{M. Fliess, C. Join, M. Mboup, H. Sira-Ram\'{\i}rez}, \guilo{}Compression
diff\'erentielle de transitoires bruit\'es\guilf{}, \newblock
\textit{C.R. Acad. Sci. Paris Ser. I}, \volumename\ 339 (2004)
821-826.

\bibitem[18]{diag}\textsc{M. Fliess, C. Join, H. Sira-Ram\'{\i}rez}, \guilo{}Residual generation
for linear fault diagnosis$:$ an algebraic setting with
examples\guilf{}, \newblock \textit{Int. J. Control}, \volumename\
77 (2004) 1223-1242.

\bibitem[19]{nl}\textsc{M. Fliess, C. Join, H. Sira-Ram\'{\i}rez}, \guilo{}Non-linear estimation
is easy\guilf{}, \newblock \textit{Int. J. Modelling Identification
Control}, \volumename\ 4 (2008) 12-27.


\bibitem[20]{fmmsr}\textsc{M. Fliess, M. Mboup, H. Mounier, H. Sira-Ram\'{\i}rez}, \guilo{}Questioning
some paradigms of signal processing via concrete examples\guilf{},
{\it in} H. Sira-Ram\'{\i}rez, G. Silva-Navarro (Eds.)$:$
\textit{Algebraic Methods in Flatness, Signal Processing and State
Estimation}, pp. 1-21, Editiorial Lagares, 2003 (accessible sur {\tt
http$:$//hal.inria.fr/inria-00001059/en/}).

\bibitem[21]{esaim}\textsc{M. Fliess, H. Sira-Ram\'{\i}rez}, \guilo{}An algebraic
framework for linear identification\guilf{}, \newblock \textit{ESAIM
Control Optim. Calc. Variat.}, \volumename\  9 (2003) 151-168.

\bibitem[22]{recons}\textsc{M. Fliess, H. Sira-Ram\'{\i}rez}, \guilo{}Reconstructeurs
d'\'{e}tat\guilf{}, \newblock \textit{C.R. Acad. Sci. Paris Ser. I},
\volumename\ 338 (2004) 91-96.

\bibitem[23]{garnier}\textsc{M. Fliess, H. Sira-Ram\'{\i}rez}, \guilo{}Closed-loop parametric
identification for continuous-time linear systems via new algebraic
techniques\guilf{}, {\it in} H. Garnier, L. Wang (Eds.)$:$ {\em
Identification of Continuous-time Models from Sampled Data}, pp.
363-391, Springer, 2008.

\bibitem[24]{green}\textsc{H.S. Green}, \textit{Information Theory and Quantum Physics},
Springer, 2000.

\bibitem[25]{gelfand}\textsc{I.N. Guelfand, N.Y. Vilenkin},
\textit{Les distributions}, t. 4: \textit{Applications de l'analyse
harmonique} (traduit du russe), Dunod, 1967.

\bibitem[26]{hida}\textsc{T. Hida, H.-H. Kuo, J. Potthof, L. Streit}, \textit{White Noise$:$
An Infinite Dimensional Calculus}, Kluwer, 1993.

\bibitem[27]{join}\textsc{C. Join, S. Tabbone}, \guilo{}Robust curvature extrema detection
based on new numerical derivation\guilf{},  \textit{Advanced
Concepts Intelligent Vision Systems -- ACIVS'2008}, Juan-les-Pins,
2008 (accessible sur {\tt
http$:$//hal.inria.fr/inria-00300799/en/}).

\bibitem[28]{glavieux}\textsc{M. Joindot, A. Glavieux}, \textit{Introduction aux
communications num\'{e}riques}, Masson, 1996.

\bibitem[29]{lang}\textsc{S. Lang}, {\it Algebra} (3$^{rd}$ rev. ed.), Springer, 2002.
Traduction fran\c{c}aise: {\it Alg\`{e}bre}, Dunod, 2004.


\bibitem[30]{liu}\textsc{D. Liu, O. Gibaru, W. Perruquetti, M. Fliess,
M. Mboup}, \guilo{}An error analysis in the algebraic estimation of
a noisy sinusoidal signal\guilf{}, \textit{Proc. 15$^{th}$ Medit.
Conf. Control Automation -- MED'2008}, Ajaccio, 2008 (accessible sur
{\tt http$:$//hal.inria.fr/inria-00300234/en/}).

\bibitem[31]{N}\textsc{C. Lobry}, {\it Et pourtant ils ne remplissent
pas $\mathbb{N}$ !}, ALEAS, 1989.

\bibitem[32]{lobry}\textsc{C. Lobry, T. Sari}, \guilo{}Nonstandard analysis and representation
of reality\guilf{}, \textit{Int. J. Control}, \volumename\ 81 (2008)
517-534.

\bibitem[33]{mboup}\textsc{M. Mboup}, \guilo{}Parameter estimation via differential
algebra and operational calculus\guilf{}, Manuscrit, 2006
(accessible sur {\tt http$:$//hal.inria.fr/inria-00138294/en/}).


\bibitem[34]{ajaccio}\textsc{M. Mboup, C. Join, M. Fliess}, \guilo{}A delay estimation approach
to change-point detection\guilf{}, \textit{Proc. 16$^{th}$ Medit.
Conf. Control Automation -- MED'2008}, Ajaccio, 2008 (accessible sur
{\tt http$:$//hal.inria.fr/inria-00179775/en/}).

\bibitem[35]{ath}\textsc{M. Mboup, C. Join, M. Fliess}, \guilo{}Numerical
differentiation with annihilators in noisy environment\guilf{},
\textit{Numer. Algorithm.}, (2009) DOI$:$ 10.1007/s11075-008-9236-1.


\bibitem[36]{McC}\textsc{J. McConnell,  J. Robson},  \textit{Noncommutative Noetherian Rings},
Amer. Math. Soc., 2000.

\bibitem[37]{nelson}\textsc{E. Nelson}, \textit{Radically Elementary Probability Theory}, Princeton
University Press, 1987 (accessible sur {\tt
http$:$//www.math.princeton.edu/\%7Enelson/books/rept.pdf}).

\bibitem[38]{neves}\textsc{A. Neves, M.D. Miranda, M. Mboup}, \guilo{}Algebraic parameter
estimation of damped exponentials\guilf{}, \textit{Proc. 15$^{th}$
Europ. Signal Processing Conf. -- EUSIPCO'2007}, Pozna\'{n}, 2007
(accesible sur {\tt http$:$//hal.inria.fr/inria-00179732/en/}).

\bibitem[39]{proakis0}\textsc{J.G. Proakis}, \textit{Digital Communications} (4$^{th}$ ed.),
McGraw-Hill, 2001.


\bibitem[40]{proakis}\textsc{J.G. Proakis, M. Salehi}, \textit{Communication Systems
Engineering} (2$^{nd}$ ed.), Prentice Hall, 2002.


\bibitem[41]{singer}\textsc{M. van der Put, M.F. Singer}, \textit{Galois Theory of Linear
Differential Equations}, Springer, 2003.

\bibitem[42]{robinson}\textsc{A. Robinson}, \textit{Non-Standard Analysis} ($2^{nd}$ ed.),
North-Holland, 1974.

\bibitem[43]{bell}\textsc{C.E. Shannon}, \guilo{}A mathematical theory of communication\guilf{},
\textit{Bell Syst. Tech. J.}, \volumename\ 27 (1948) 379-457 \&
623-656.

\bibitem[44]{ire}\textsc{C.E. Shannon}, \guilo{}Communication in the presence of noise\guilf{},
\textit{Proc. IRE}, \volumename\ 37 (1949) 10-21.

\bibitem[45]{trapero}\textsc{J.R. Trapero, H. Sira-Ram\'{\i}rez, V. Feliu Battle},
\guilo{}An algebraic frequency estimator for a biased and noisy
sinusoidal signal\guilf{}, \textit{Signal Processing}, \volumename\
87 (2007) 1188-1201.

\bibitem[46]{trapero-bis}\textsc{J.R. Trapero, H. Sira-Ram\'{\i}rez, V. Feliu Battle}, \guilo{}A fast
on-line frequency estimator of lightly damped vibrations in flexible
structures\guilf{}, \textit{J. Sound Vibration}, \volumename\ 307
(2007) 365-378.


\bibitem[47]{trapero-ter}\textsc{J.R. Trapero, H. Sira-Ram\'{\i}rez, V. Feliu Batlle},
\guilo{}On the algebraic identification of the frequencies,
amplitudes and phases of two sinusoidal signals from their noisy
sums\guilf{}, \textit{Int. J. Control}, \volumename\ 81 (2008)
505-516.

\bibitem[48]{yosida}\textsc{K. Yosida}, \textit{Operational Calculus$:$ A Theory of
Hyperfunctions}  (translated from the Japanese), Springer, 1984.


\end{thebibliography}
\end{document}